\documentclass[9pt]{article}

\usepackage{spconf,amsmath,graphicx,hyperref}
\DeclareMathOperator{\RMSE}{RMSE}

\usepackage{cite}
\usepackage{amssymb,amsfonts}
\usepackage{algorithmic}
\usepackage{textcomp}

\usepackage{siunitx} 
\usepackage{bm}
\usepackage{booktabs}
\usepackage{color}
\usepackage{url}
\usepackage{multirow}
\usepackage{diagbox}
\usepackage{float}
\usepackage{subfig}
\usepackage{enumitem}
\usepackage{xurl}
\usepackage{mathtools}

\usepackage[table]{xcolor} % Include xcolor package with table option
\definecolor{mycolor}{RGB}{230, 230, 255} % Define a custom color

\usepackage{etoolbox}
\usepackage{array}
\apptocmd{\thebibliography}{\setlength{\itemsep}{0.9pt}\setlength{\parskip}{0.8pt}\linespread{0.9}\selectfont}{}{}

\title{Attentive AV-FusionNet: \\ Audio-Visual Quality Prediction with Hybrid Attention}
\name{Ina Salaj* and Arijit Biswas\thanks{*Research performed as a student at TH-Ingolstadt and as an intern at Dolby Germany GmbH.}}
\address{Dolby Germany GmbH, Nürnberg, Germany}
\begin{document}
\maketitle
\begin{abstract}
We introduce a novel deep learning–based audio-visual quality (AVQ) prediction model that leverages internal features from state-of-the-art unimodal predictors. Unlike prior approaches that rely on simple fusion strategies, our model employs a hybrid representation that combines learned Generative Machine Listener (GML) audio features with hand-crafted Video Multimethod Assessment Fusion (VMAF) video features. Attention mechanisms capture cross-modal interactions and intra-modal relationships, yielding context-aware quality representations. A modality relevance estimator quantifies each modality's contribution per content, potentially enabling adaptive bitrate allocation. Experiments demonstrate improved AVQ prediction accuracy and robustness across diverse content types.
\end{abstract}

\section{Introduction}
\label{section:Introduction}

The rapid growth of multimedia streaming has made optimizing user-perceived audio and video quality under bandwidth constraints a critical challenge. Streaming platforms must balance compression efficiency with perceptual Quality of Experience (QoE), which depends on both signal integrity and human perception; impairments in one modality can degrade the overall experience even if the other remains intact~\cite{qoe, multimediaQoE}. While objective Quality of Service (QoS) metrics, such as bitrate and latency, are easily measured, they often fail to predict perceptual quality~\cite{qoeApplication}. Subjective evaluations remain the gold standard~\cite{subAssessment, itu910} but are costly and impractical at scale, motivating the development of objective AVQ metrics.

Early AVQ research highlighted strong cross-modal effects, where video quality can influence perceived audio quality and vice versa~\cite{hollier1997objective, beerends1999the}. However, many existing models were developed with outdated codecs and less sophisticated unimodal predictors. Datasets such as LIVE-SJTU~\cite{Bovik_AVQ} and UnB-AV~\cite{martinez2018immersive} have restricted degradations, limiting their relevance for modern streaming. Meanwhile, unimodal quality assessment has advanced considerably: video metrics evolved from PSNR and SSIM~\cite{imageQA} to perceptually optimized models such as VMAF~\cite{NetflixTechBlogVMAF}, while audio models progressed from PEAQ~\cite{peaqFirst} and ViSQOL-v3~\cite{v3} to deep learning–based predictors such as GML~\cite{gml}. Extending these advances to joint audio–visual content remains challenging, particularly in capturing cross-modal interactions and long-range dependencies.

Existing AVQ approaches often oversimplify modality fusion. Late-fusion methods~\cite{basicQM, AVmetrics, Bovik_AVQ} ignore dynamic dependencies, such as when strong audio masks visual flaws or high-quality video compensates for audio distortions~\cite{visualInfluence, crossmodaleffects}. Moreover, the relative importance of audio and video varies by content: conversational scenes tolerate lower video quality, while action sequences require higher visual fidelity~\cite{visionmodels}. Static fusion fails to capture content-dependent effects, motivating models that adaptively fuse audio and video based on their relative importance.

In this work, we revisit AVQ modeling from a full-reference (FR) perspective because it allows systematic study. Among existing deep learning-based FR AVQ models~\cite{Bovik_AVQ, AttAVQ_Cao}, the former (Model Type-4 in~\cite{Bovik_AVQ}) does not model cross-modal interaction. The latter~\cite{AttAVQ_Cao}, on the other hand, employs an audio quality metric that may not fully reflect the current state of the art and also does not explicitly model cross-modal interactions. Motivated by these gaps, we propose a full-reference AVQ model, Attentive AV-FusionNet, that leverages cross- and self-attention to capture both intra- and inter-modal dependencies. Features from state-of-the-art unimodal predictors (GML for audio, VMAF for video) are aligned and fused via bidirectional cross-attention, followed by self-attention to refine context-aware representations. A modality relevance estimator quantifies the contribution of each modality per content, potentially enabling adaptive bitrate allocation.

Main contributions:
\begin{itemize}[topsep=0pt, parsep=0pt, itemsep=0pt]
\item A hybrid AVQ model integrating learned GML and hand-crafted VMAF features, with attention-based fusion to capture cross-modal interactions.
\item A modality relevance estimator that provides content-aware insights.
\item Extensive evaluation on internal and external datasets, demonstrating improved prediction accuracy and robustness.
\end{itemize}

The remainder of this paper describes our proposed model (Section~\ref{section:model}), modality importance estimation (Section~\ref{section:modality-imp}), datasets (Section~\ref{section:dataset}), and experimental results (Section~\ref{sec:results}) before concluding (Section~\ref{section:conclusion}).

\begin{figure*}[t]
    \centering
    \includegraphics[width=0.94\linewidth]{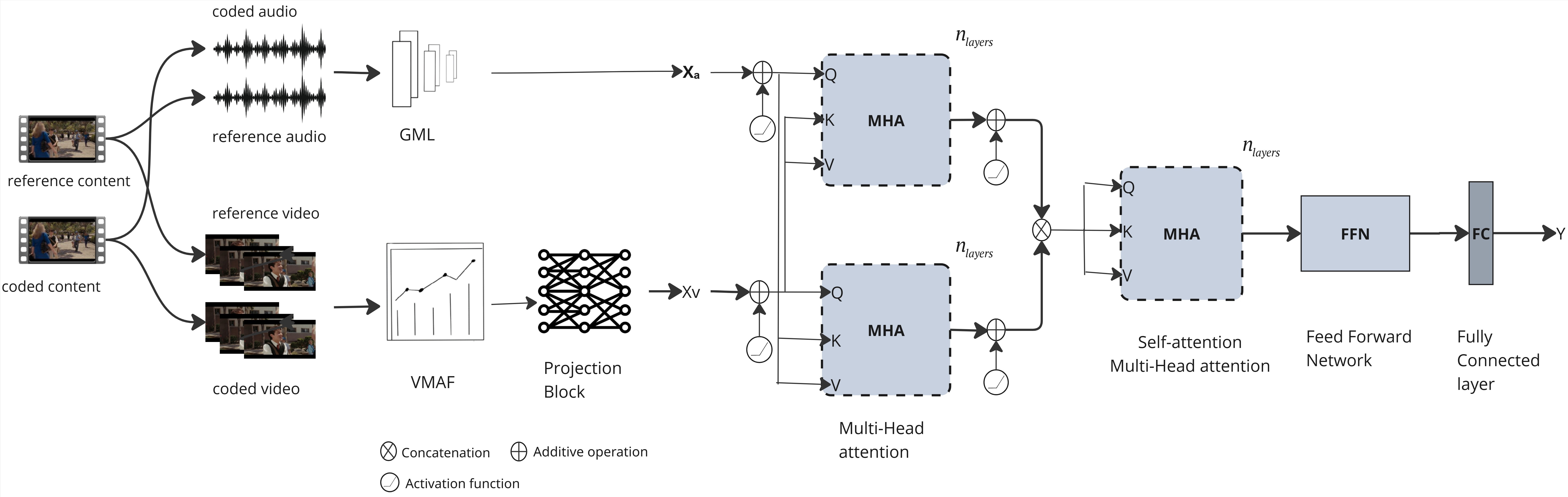} 
    \caption{Attentive AV-FusionNet: joint audio–visual quality prediction model integrating GML~\cite{gml} and VMAF~\cite{NetflixTechBlogVMAF} features, with 7.4M trainable parameters in the projection and fusion network.}
    \label{fig:AttAVfusionNet}
\end{figure*}

% \begin{itemize}[topsep=0pt, parsep=0pt, itemsep=0pt]
% \item We design a full-reference audio–visual quality prediction model that leverages modality-specific quality estimators (GML for audio and VMAF for video) and integrates them through a hybrid attention-based fusion mechanism. This design enables the model to capture both intra-modal context and inter-modal dependencies, resulting in refined and context-aware quality reasoning.  
% \item We introduce a modality relevance estimator that quantifies the relative contribution of audio and video on a per-content basis. This would allow bitrate-aware decision making and support more efficient content adaptation.  
% \end{itemize}

\section{AVQ Model: Attentive AV-FusionNet}
\label{section:model}

\textit{Attentive AV-FusionNet} has three stages: feature extraction, attention-based fusion, and prediction.

\textbf{Feature Extraction.}  
Video features are extracted from VMAF's~\cite{NetflixTechBlogVMAF} internal representations, aggregating frame-level data to encode spatiotemporal information. We use VMAF as a robust video quality predictor, following~\cite{Wien2024}. Features include four Visual Information Fidelity (VIF)~\cite{VIFP} features, one motion (Motion2)~\cite{NetflixTechBlogVMAF}, and one Additive Detail Metric (ADM)~\cite{DLM}, resulting in $X_v \in \mathbb{R}^{N \times d_v}$ with $d_v=6$ and $N$ temporal units after pooling. In our implementation, $N{=}1$, i.e., clip-level.

Audio features are obtained from deeper layers of the GML~\cite{gml}, just before the final fully connected layer yielding $X_a \in \mathbb{R}^{N \times d_a}$ with $d_a=512$.

To align modalities, video features are projected into the audio space via a learnable mapping:
\begin{equation}
X_v' = \sigma(X_v W_v), \quad W_v \in \mathbb{R}^{d_v \times 512}.
\end{equation}
where \(\sigma(\cdot)\) is a non-linear activation (e.g., GELU). This expansion preserves audio discriminability while keeping the module lightweight, placing both modalities in a 512-dimensional space for symmetric cross-modal mixing. %With both modalities aligned in dimensionality after projection ($d = 512$), we apply cross-attention to explicitly model their interactions.

\textbf{Attention-based Fusion.}  
Fusion uses bidirectional multi-head cross-attention, following the standard Transformer formulation~\cite{Attention}. In the audio-to-visual direction (single-head shown; in practice, $H$ heads with $d_k = 512/H$ are used):
\begin{equation}
\begin{gathered}
Q = X_v' W_Q,\quad K = X_a W_K,\quad V = X_a W_V,\\[-2pt]
X_v^{ca} = \mathrm{softmax}\!\left(\frac{QK^\top}{\sqrt{d_k}}\right)\,V.
\end{gathered}
\end{equation}

The reciprocal visual-to-audio case is analogous, yielding \(X_a^{ca}\). The concatenated joint representation is
$ J = [X_a^{ca}; X_v^{ca}] \in \mathbb{R}^{N \times 1024}.$
Here, the 1024-dimensional ($d_j$) representation results from concatenating the 512-dimensional cross-attended features of audio and video, effectively combining information from both modalities.

Self-attention refines the joint representation:
\begin{equation}
\begin{gathered}
Q = J W_Q',\quad K = J W_K',\quad V = J W_V',\\[-2pt]
J_{\text{self}} = \mathrm{softmax}\!\left(\frac{QK^\top}{\sqrt{d_k^{(j)}}}\right)\,V,
\end{gathered}
\end{equation}
where $d_k^{(j)} = d_j/H_j$. We omit positional encodings and residual connections, as the pooled VMAF and GML features already encode temporal structure. Preliminary experiments confirmed that adding residuals did not improve performance, and omitting them helps preserve modality-specific contributions without over-smoothing.

\textbf{Prediction.}  
The refined features \(J_{\text{self}}\) are passed through a shallow feed-forward head with non-linear activations (e.g., GELU, Tanh) and dropout:
\begin{equation}
\hat{Q}_{av} = f_{\text{FFN}}(J_{\text{self}}).
\end{equation}
producing the predicted audio-visual quality score. The head is deliberately shallow to avoid overfitting; performance gains stem primarily from the attention mechanisms.

\textbf{Training Objective.}  
We employ a loss combining the Concordance Correlation Coefficient (CCC)~\cite{lawrence1989concordance} and Root Mean Square Error ($\RMSE$):
\begin{equation}
L = \lambda \,(1 - \text{$CCC$}) + (1 - \lambda) \,\text{$RMSE$}, \quad \lambda \in [0,1].
\label{eq:ccc_rmse_loss}
\end{equation}
Empirically, \(\lambda=0.6\) balances correlation with subjective ratings while minimizing absolute error.

\section{Assessing Modality Importance}
\label{section:modality-imp}

Beyond a single AVQ score, we quantify each modality's contribution using two complementary metrics: ablation sensitivity and feature change norms.

\textbf{Ablation Sensitivity.} For modality $m \in \{\text{audio}, \text{video}\}$, this metric measures the prediction error when $m$ is masked:
\begin{equation}
    I^{\text{abl}}_m = \frac{\| f(x) - f(x \setminus m) \|_2}{\| f(x) \|_2},
\end{equation}
where $f(x)$ is the full-model output and $f(x \setminus m)$ is the output with $m$ zeroed. Higher values indicate stronger reliance. Masking video typically increases errors in action-heavy clips, whereas audio dominates in conversational or music-driven content.

\textbf{Feature Change Norm.} This metric assesses how much a modality adapts during fusion:
\begin{equation}
    I^{\text{norm}}_m = \frac{\| \phi^{\text{pre}}_m - \phi^{\text{post}}_m \|_2}{\| \phi^{\text{pre}}_m \|_2},
\end{equation}
where $\phi^{\text{pre}}_m$ and $\phi^{\text{post}}_m$ are embeddings before and after cross-attention. Smaller values indicate greater stability and independent contribution. Although the two measures capture different aspects, they remain consistent in practice, suggesting that our analysis is robust.

\begin{figure}[t]
    \centering
    \subfloat[Audio-dominant. \label{fig:modimp-a}]{
        \includegraphics[width=0.35\linewidth]{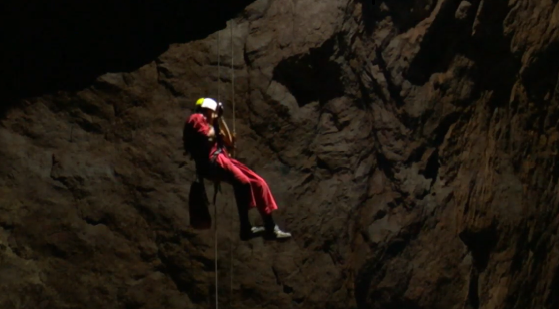}
    }\hfill
    \subfloat[Video-dominant. \label{fig:modimp-b}]{
        \includegraphics[width=0.35\linewidth]{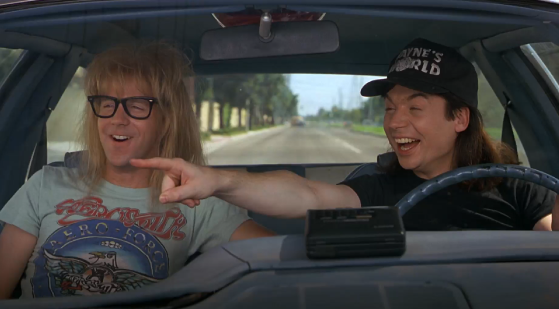}
    }
    \caption{Modality importance estimation for different content types.}
    \label{fig:modality-imp}
\end{figure}

The final importance score combines the two metrics:
\begin{equation}
I^{\text{final}}_m = \alpha \cdot \mathrm{norm}(I^{\text{abl}}_m) + \beta \cdot [1 - \mathrm{norm}\big(I^{\text{norm}}_m\big)],
\end{equation}
where $\mathrm{norm}(\cdot)$ denotes min--max normalization. We set $\alpha=0.7$ and $\beta=0.3$ to prioritize ablation sensitivity, yielding an interpretable measure of modality importance. As shown in Fig.\ref{fig:modality-imp}, the model adapts content-dependently: Fig.~\ref{fig:modality-imp}(a) shows a scene with audio containing dense transients (a known codec challenge~\cite{biswas18b_ac4companding}), whereas Fig.~\ref{fig:modality-imp}(b) depicts a visually dynamic scene with easy-to-code music. The model appears to assign higher relevance to audio when impairments are salient and to video when visual distractions may mask them. Although subjective data on modality relevance are unavailable, preventing absolute accuracy quantification, observed trends remain consistent when a modality’s bitrate is fixed and the other is varied.

\section{Datasets}
\label{section:dataset}

We train on the internal full-reference AVQ dataset~\cite{Tuncok_darcy_dolby}, which contains 65 source clips across diverse genres (e.g., films, documentaries, animations, and sports). 
Each clip was encoded at five H.264 video bitrates (0.5, 1, 2, 4, 25 Mb/s) and five stereo audio bitrates (16, 32, 64, 96, 256 kb/s), yielding \(65 \times 25 = 1625\) stimuli. 
The lowest two audio bitrates use HE-AACv2~\cite{HE-AAC}, the next two HE-AACv1~\cite{HE-AAC}, and the highest a standard AAC codec~\cite{AAC}. 
Subjective ratings were collected under ITU-compliant conditions on a 5-point MOS scale from ten participants, totaling 16,250 ratings. For experiments, we held out five source clips for a test set of 125 stimuli (25 AV combinations per clip), with the remaining 1,500 stimuli forming the training set. 

For external evaluation, we use the LIVE-SJTU AV-QA database~\cite{Bovik_AVQ}, a widely adopted benchmark for multimodal quality assessment. 
It contains 14 source videos spanning cartoons, indoor/outdoor real-life scenes, concerts, documentaries, theater, and sports. 
Videos were encoded with H.265 at multiple constant rate factor (CRF) settings, and stereo audio with AAC spanning 8–128 kb/s. 
Subjective ratings were collected using SSCQE~\cite{series2012methodology} and aggregated to MOS via Z\textendash score normalization, scaled to \([0,100]\). 
To reflect practically relevant scenarios, we restrict evaluation to AAC 32–128 kb/s, omitting 8 kb/s.
%Extremely low-rate audio (8 kb/s) was excluded, as it does not represent practical scenarios.

\begin{table*}[ht]
\centering
\fontsize{9}{10}\selectfont  % smaller font
\setlength{\tabcolsep}{4pt}  % tighter horizontal spacing
\renewcommand{\arraystretch}{0.9}  % tighter row spacing
\caption{Comparison of model performance on internal and external (LIVE-SJTU) datasets. Metrics: Pearson ($R_p$), Spearman ($R_s$), and $\RMSE$. Best values per column within each category are \textbf{bold}, with the top model highlighted.}
\label{tab:all_models}
\begin{tabular}{l l c c c c c c}
\toprule
\multirow{2}{*}{Category} & \multirow{2}{*}{Model} & \multicolumn{3}{c}{Internal Dataset} & \multicolumn{3}{c}{LIVE-SJTU} \\
\cmidrule(lr){3-5} \cmidrule(lr){6-8}
 & & $R_p\uparrow$ & $R_s\uparrow$ & RMSE$\downarrow$ & $R_p\uparrow$ & $R_s\uparrow$ & RMSE$\downarrow$ \\
\midrule
Baseline & Q-Random (best guess $w_a=0.3$, $w_v=0.7$)  & \textbf{0.84} & 0.86 & 1.13 & 0.85 & 0.85 & 0.98 \\
 & Q-Internal (optimal $w_a=0.33$, $w_v=0.67$) & 0.83 & 0.86 & \textbf{1.12} & 0.82 & 0.82 & 1.04 \\
 \rowcolor{blue!05} 
 & Q-External (optimal $w_a=0.23$, $w_v=0.76$) & 0.83 & \textbf{0.87} & 1.13 & \textbf{0.90} & \textbf{0.91} & \textbf{0.68} \\
\midrule
SVR & SVR-2F (Q$_a$, Q$_v$) & 0.86 & 0.84 & 0.53 & 0.73 & 0.78 & 0.94 \\
 & SVR-3F (Q$_a$, Q$_v$, audio bitrate)   & 0.89 & 0.88 & 0.47 & 0.89 & 0.91 & \textbf{0.47} \\
 & SVR-7F (Q$_a$, 6 VMAF features) & 0.86 & 0.86 & 0.50 & 0.88 & 0.92 & 0.84 \\
 \rowcolor{blue!10} 
 & SVR-8F (Q$_a$, 6 VMAF features, audio bitrate)   & \textbf{0.90} & \textbf{0.89} & \textbf{0.43} & \textbf{0.90} & \textbf{0.95} & 0.86 \\
\midrule
Deep Learning & Simple AV-Fusion (ours without attention)      & 0.84 & 0.83 & 0.62 & 0.89 & 0.89 & 1.06 \\
 & CA AV-Fusion (ours without self-attention)         & 0.90 & 0.87 & 0.47 & 0.90 & 0.91 & 0.59 \\
 & Recursive AV-FusionNet & 0.90 & 0.89 & 0.47 & \textbf{0.92} & 0.91 & \textbf{0.39} \\
 \rowcolor{blue!15} 
 & Attentive AV-FusionNet (ours) & \textbf{0.97} & \textbf{0.96} & \textbf{0.22} & \textbf{0.92} & \textbf{0.92} & 0.44 \\
\bottomrule
\end{tabular}
\end{table*}

\section{Experiments and Results}
\label{sec:results}

To ensure comparability, we first aligned the datasets' subjective quality scales. The internal dataset uses a 5-point MOS scale, while LIVE-SJTU uses a continuous $[0,100]$ scale. Scores were rescaled with the IBM transformation~\cite{ibm_likert_scales} to preserve relative distributions.
Support Vector Regression (SVR) employed feature normalization to the $[0,1]$ range, while deep learning models used per-feature standardization over the entire dataset. The external dataset (LIVE-SJTU) was used exclusively for testing to evaluate cross-dataset generalization. Performance was assessed using Pearson correlation ($R_p$), Spearman correlation ($R_s$), and $\RMSE$. High correlations indicate alignment with subjective judgments, while low $\RMSE$ reflects predictive accuracy.

\subsection{Experimental Comparison}

\textbf{Weighted Product Model (Baseline).} We implemented a weighted product model to fuse unimodal quality scores $Q_a$ and $Q_v$, where $Q_a$ denotes the audio quality prediction from GML, and $Q_v$ denotes the video quality prediction obtained using VMAF:
\begin{equation}
Q_{av} = Q_a^{w_a} \cdot Q_v^{w_v},
\end{equation}
with $w_a$ and $w_v$ representing the modality weights. We tested three strategies: random (Q-Random) best guess, one optimized on the internal dataset (Q-Internal), and one on the external dataset (Q-External). Weights were optimized by minimizing mean squared error between predicted and ground-truth scores.
Results (Table~\ref{tab:all_models}) indicate that simple multiplicative fusion is insufficient for modeling complex cross-modal interactions, providing a lower performance bound for subsequent methods.

\textbf{Support Vector Regression (SVR).} SVR was applied as a regression-based fusion method with Radial Basis Function kernels and multiple feature configurations. Three feature families were tested: (i) $Q_a$ and $Q_v$ only (SVR-2F), (ii) $Q_a$ and $Q_v$ plus audio bitrate (SVR-3F), and (iii) extended video features derived from VMAF. Within the latter, two variants were considered: SVR-7F, which includes $Q_a$ together with six VMAF features, and SVR-8F, which additionally includes audio bitrate. Video bitrate was intentionally excluded, since the internal dataset provides constant bitrates (CBR), while the external dataset is encoded using a CRF, leading to incompatible feature alignment.
On LIVE-SJTU, SVR achieves correlations up to 0.95 and $\RMSE$ as low as 0.47. Performance is lower on the internal dataset, where SVR-8F reaches $R_p=0.90$ and $\RMSE=0.43$. Cross-dataset evaluation indicates that models trained on the internal dataset generalize well to LIVE-SJTU, whereas the reverse is less effective, suggesting that LIVE-SJTU has simpler degradation patterns.

\textbf{Deep Learning Models.} To capture nonlinear cross-modal interactions and long-range dependencies, we evaluated several attention-based neural architectures, tuning hyperparameters via validation (Table~\ref{tab:hyperparameters}). We started with a Simple AV-Fusion baseline, which is our proposed model without any attention mechanism. We then added cross-attention to create CA AV-Fusion, a variant that explicitly models inter-modal dependencies. We also explored Recursive AV-FusionNet, an alternative architecture inspired by~\cite{Praveen_EmotionRecog}. This variant applies cross-attention iteratively to a joint audio-visual representation, where each modality attends to a shared embedding. This allowed us to specifically test whether modalities benefit more from attending to a unified multimodal context rather than engaging in direct, bidirectional interaction. Our final proposed model, Attentive AV-FusionNet, integrates both cross- and self-attention. This design not only models the direct interactions between audio and video but also refines each modality's context-aware representation, leading to more accurate and robust quality predictions (Table~\ref{tab:all_models}).
We train with 5-fold cross-validation (batch size = 32) using AdamW \cite{kingma2017adam}; the search space and selected settings are given in Table~\ref{tab:hyperparameters}.

%Extensive hyperparameter searches (Table~\ref{tab:hyperparameters}) showed that $d_{model}=512$, GELU activations, and moderate dropout yield the best trade-off between accuracy and generalization. 

\begin{table}[ht]
\centering
\footnotesize
\caption{Hyper-parameter search space. \textbf{Best configuration} selected by validation loss; dimensions marked \textsuperscript{*} are fixed by design (6→512 projection; 512+512→1024 concatenation).}
\label{tab:hyperparameters}
\setlength{\tabcolsep}{2pt} % reduce horizontal padding
% +0.005 to the tight columns, -0.005 from their partners
\newcolumntype{V}[1]{>{\hspace*{-2pt}\raggedright\arraybackslash}p{#1}} % was -6pt
\begin{tabular}{%
  p{0.225\columnwidth} p{0.245\columnwidth} |
  p{0.225\columnwidth} p{0.225\columnwidth}}

\hline
\rowcolor{blue!13}
\textbf{Hyper-parameter} & \textbf{Value Range} & \textbf{Hyper-parameter} & \textbf{Value Range} \\
\hline
\rowcolor{blue!5} \multicolumn{4}{l}{\textit{General Parameters}} \\
$n_{layers}$       & [1,\textbf{2},...,5]          & $\lambda$          & [0.0,\textbf{0.6},..,1.0] \\
activation         & \textbf{GELU}, ReLU      & dropout           & [0.1, ..., \textbf{0.6}] \\
learning rate  & $10^{\{\mathbf{-4},\ldots,-2\}}$  & weight decay & $10^{\{\mathbf{-3},\ldots,-1\}}$  \\
\rowcolor{blue!5} \multicolumn{4}{l}{\textit{Attention Models}} \\
$d_{\text{model}}^{\text{cross}}$ & \textbf{512}\textsuperscript{*} & $d_{\text{model}}^{\text{joint}}$ & \textbf{1024}\textsuperscript{*} \\
$n_{\text{heads}}\textit(H)$ & $\{2,\textbf{4},8\}$ & $d_{\text{feedforward}}$ & $\{256,\textbf{512},1024\}$ \\
\rowcolor{blue!5} \multicolumn{4}{l}{\textit{Cross-modal Attention}} \\
$n_{layers}^{V \rightarrow A}$ & [\textbf{1},5] & $n_{layers}^{A \rightarrow V}$ & [\textbf{1},5] \\
\hline
\end{tabular}
\end{table}

\section{Conclusion}
\label{section:conclusion}

We proposed Attentive AV-FusionNet, a full-reference AVQ prediction model that fuses GML audio and VMAF video features through hybrid attention. By explicitly modeling cross- and intra-modal dependencies, the model achieves higher accuracy and robustness than baseline and SVR methods on both internal and external datasets. In addition to improved predictive performance, it provides interpretable estimates of modality relevance, which could enable content- and bitrate-aware adaptation. Future work will extend the approach to real-time scenarios, more diverse content, explainable attention analysis, and the collection of subjective data to better quantify modality relevance.

\bibliographystyle{IEEEbib}
\bibliography{strings,refs}

\end{document}